\documentclass[technote,12pt,onecolumn]{IEEEtran}
\usepackage{amsmath}
\usepackage{amssymb}
\usepackage[dvips]{graphicx}
\begin{document}
\title{On the calculation of Schottky contact resistivity}
\author{Yang Liu \\
Center for Integrated Systems, \\
Stanford University, Stanford, CA 94305-4075 \\
E-mail: yangliu@gloworm.stanford.edu }
\maketitle
\section*{Abstract}
This numerical study examines the importance of self-consistently accounting for transport and electrostatics in the calculaiton
of semiconductor/metal Schottky contact resistivity. It is shown that ignoring such self-consistency results in significant
under-estimation of the contact resistivity. An explicit numerical method has also been proposed to efficiently improve
contact resistivity calculations.

\section{Introduction}
In modern MOSFET designs, silicon/silicide Schottky hetero-interfaces are commonly used to form source/drain contacts.
As the device size continues to shrink with each generation, the resistance associated with the silicon/silicide
contacts begins to have its impact on the device performance~\cite{Hu}. Therefore, it becomes important to achieve
an accurate yet efficient method to evaluate silicon/silicide contact resistivity at different doping concentrations
and temperatures. Some widely refered earlier work on J-V relation and contact resistance of Schottky contact are 
due to~\cite{Padovani,Crowell,Yu}. They identified three regimes where the major current contribution
comes from field emission, thermionic field emission or thermionic emission, depending on doping concentration and temperature. 
Analytical expressions of contact resistivity were obtained for each of the three regimes. However, their works
were based on two assumptions. Firstly, Poisson equation was only solved by assuming fully depletion of free carriers within the
depletion region, i.e. the band-bending of silicon near the interface is parabolic. Secondly, the continuity equation was not
solved, and therefore the carrier quasi-Fermi level was assumed to be flat across the depletion region. 
To examine the validity of those two
assumptions, we have implemented in our device simulator Prophet a self-consistent Schottky barrier diode simulation model 
proposed by~\cite{Matsuzawa}. In this physical model, Poisson and continuity equations are simultaneously solved with the distributed
tunneling current self-consistently included. By comparing analytical models against this physical model, 
we are able to demonstrate that those two assumptions lead to significant deviation in contact 
resistivity calculations, particularly at high doping concentration of modern device and high temperature under ESD condition. 
In this work, we also extend the analytical model by removing those two assumptions. In the improved model, 
a unified contact resistance expression is explicitly obtained\footnote{Numerical integrations are involved in 
the expression.}, and the results show excellent match with 
those from the physical model. 

\section{Models and Results}
In Fig.1 is a schematic energy band-diagram of a silicon/silicide Schottky contact. 
In this work, we assume the majority carriers are
electrons. The hetero-interface in this plot is at location $w$
and the depletion region is between $0$ and $w$. The difference of silicon/silicide affinity is $q\phi_b$ where $q$ is the
elementary charge. The doping density in silicon region is $N_D$ and the effective density of states is $N_c$. The difference
between conduction band and quasi-fermi level at charge neutral region is $q\phi_s$. The applied forward bias is $V_f$. 
If we only consider the contact resistance at very low
bias, we have $V_f\rightarrow 0$. The barrier height $E_b$ can be expressed as $E_b=q(\phi_b-\phi_s-V_f)$. We base this work
on Maxwell-Boltzmann statistics for mathematical simplicity~\footnote{Also because the physical model is currently 
implemented based on M-B statistics.}. All the derivations below can
be readily extended to Fermi-Dirac statistics, and the effects investigated in this work should still be present in that case.
It can be shown that $exp(-q\phi_s/kT)=N_D/N_c$ under M-B statistics.

Following the treatment of~\cite{Bardeen, Harrison},
the general tunneling probability at loaction $x$ within the depletion region is given by
\begin{eqnarray}
\tau(E(x))&=&exp\Bigl(-\frac{4\pi}{h}\int_x^w\sqrt{2m^*[E(x')-E(x)]}dx'\Bigr) \nonumber \\
&=&exp\Bigl(-\frac{4\pi}{h}\int_{E(x)}^{E_b}\sqrt{2m^*[E(x')-E(x)]}\frac{dx'}{dE(x')}dx'\Bigr)
\label{eq:tau_E}
\end{eqnarray}
where $m^*$ is the tunneling effective mass, 
$E(x)$ is the conduction bandedge energy at position $x$. The electro-static potential $\phi(x)$ is given
by $\phi(x)=-E(x)/q$. It can be seen that, if $dx/dE(x)$ is known, the tunneling probability $\tau(E)$ can be
evaluated explicitly by performing the numerical integral in Eqn. 1. Since the bandedge energy $E(x)$ is a monotonic function
of position $x$, it is convenient to use a dimensionless quantity $\alpha\equiv E(x)/E_b$ as the basic variable~\cite{Crowell}.
 In doing so, the general expression for the tunneling probability is re-written as
\begin{equation}
\tau(\alpha)=exp\Bigl(-\frac{E_b}{E_{00}}\int_{\alpha}^{1}\sqrt{\frac{\alpha'-\alpha}{F(\alpha')}}d\alpha'\Bigr),
\label{eq:tau_alpha}
\end{equation}
where $E_{00}\equiv (qh/4\pi)\sqrt{N_D/m^*\epsilon}$. The function $F(\alpha)$ carries the information of potential
variation in the depletion region and is defined as
\begin{equation}
F(\alpha)\equiv \frac{\epsilon}{2q^2N_DE_b}\bigl(\frac{dE}{dx}\bigr)^2=\frac{\epsilon E_b}{2q^2N_D}
\bigl(\frac{d\alpha}{dx}\bigr)^2.
\label{eq:F_alpha}
\end{equation}

In the literatures, different energy band-diagrams ($E(x)$) have been assumed to calculate $\tau(\alpha)$. In 
the work of~\cite{Crowell}, they based their calculation completely on the parabolic band-bending relation
\begin{equation}
E(x)=qN_Dx^2/(2\epsilon)
\label{eq:E_x}
\end{equation}
and obtained an analytical expression
\begin{equation}
\tau^{C.R.}(\alpha)=exp[-\frac{E_b}{E_{00}}y(\alpha)],
\label{eq:tau_CR}
\end{equation}
where $y(\alpha)\equiv \sqrt{1-\alpha}-\alpha log[(1+\sqrt{1-\alpha})/\sqrt{\alpha}]$.
However, the work in~\cite{Padovani,Matsuzawa} was based on another widely used formular for tunneling probability:
\begin{equation}
\tau^{P.S.}(E)=exp\Bigl[-\frac{8\pi}{3}\frac{\sqrt{2m^*}}{h}\frac{(E_b-E)^{3/2}}{|\mathbf{F}|}\Bigr],
\end{equation}
where $\mathbf{F}$ is the electrical field perpendicular to the interface. Matsuzawa et al. used 
$|\mathbf{F(x)}|=(E_b-E(x))/(w-x)$ in their work~\cite{Matsuzawa}, and therefore the tunneling probability can be re-written as
\begin{equation}
\tau^{P.S.}(E)=exp\Bigl[-\frac{8\pi}{3}\frac{\sqrt{2m^*}}{h}\sqrt{E_b-E}\cdot(w-x)\bigr].
\label{eq:tau_PS}
\end{equation}
This expression can also be directly derived from the general expression (Eqn.~\ref{eq:tau_E}) by assuming
linear relation between $E$ and $x$. It should be noted that although triangular potential barrier was assumed
in obtaining Eqn.~\ref{eq:tau_PS}, the position $x$ appears explicitly in the formula, which still contains
the band-bending information. In our physical model, the treatment follows that of~\cite{Matsuzawa}, and Eqn.~\ref{eq:tau_PS} is
adopted. In the simulation, the realistic potential variation enters through the relation of $x(E)$ by solving 
the Poisson and continuity equations. In order to have fair comparisons between numerical models and the physical
model, we use Eqn.~\ref{eq:tau_PS} for the numerical models in most cases throughout this work.

The general expression for the forward current density (electrons inject from silicon to silicide) is given by
\begin{equation}
J_f=\frac{A}{kT}e^{-q\phi_s/kT}\Bigl[\int_{0}^{E_b}\tau(E)e^{-\frac{E+q\eta}{kT}}dE+e^{-\frac{E_b+q\eta_b}{kT}}\Bigr],
\label{eq:Jf}
\end{equation}
where $A\equiv A^*T^2$ for Richardson constant $A^*\equiv 4\pi m^*qk^2/h^3$, $\eta$ is the local electron quasi-Fermi level
variation compared with that at charge neutral region. Therefore, if constant quasi-Fermi level is assumed, $\eta$ is always
zero within the entire silicon region. The first term
in the bracket of Eqn.~\ref{eq:Jf} corresponds to field emission and thermionic field emission,
while the second term is from the thermionic emission.
Similarly, the reverse current density (electron inject from silicide to silicon) is given by
\begin{equation}
J_r=\frac{A}{kT}e^{-q\phi_s/kT}\Bigl[\int_{0}^{E_b}\tau(E)e^{-\frac{E+qV_f}{kT}}dE+e^{-\frac{E_b+qV_f}{kT}}\Bigr].
\label{eq:Jr}
\end{equation}
The total net current density as a function of $V_f$ is therefore obtained as
\begin{eqnarray}
J &\equiv& J_f-J_r \nonumber \\
&=&\frac{A}{kT}e^{-q\phi_s/kT}\Bigl[\int_{0}^{E_b}\tau(E)e^{-\frac{E}{kT}}
\bigl(e^{-\frac{q\eta}{kT}}-e^{-\frac{qV_f}{kT}}\bigr)dE
+e^{-\frac{E_b}{kT}}\bigl(e^{-\frac{q\eta_b}{kT}}-e^{-\frac{qV_f}{kT}}\bigr)\Bigr],
\label{eq:J_tot}
\end{eqnarray}
where $\eta_b$ is the value of $\eta$ at the interface.
The contact resistivity at zero bias is then obtained as 
$R\equiv (dJ/dV_f)^{-1}|_{V_f\rightarrow 0}$~\cite{Yu}. In previous work~\cite{Padovani,Crowell}, continuity equation was not solved. Therefore, the variation of $\eta$ 
within the depletion region was
ignored. Under such an assumption, the total current density is obtained as
\begin{equation}
J=\frac{A}{kT}e^{-q\phi_s/kT}\Bigl[\int_{0}^{E_b}\tau(E)e^{-\frac{E}{kT}}dE+e^{-\frac{E_b}{kT}}\Bigr]
\Bigl[1-e^{-\frac{qV_f}{kT}}\Bigr].
\label{eq:J_tot_const}
\end{equation}
Hence, the inverse of contact resistivity at zero bias is
\begin{eqnarray}
R^{-1}&=&\frac{Aq}{(kT)^2}e^{-q\phi_s/kT}\Bigl[\int_{0}^{E_b}\tau(E)e^{-\frac{E}{kT}}dE+e^{-\frac{E_b}{kT}}\Bigr] \nonumber \\
&=&\frac{AqE_b}{(kT)^2}e^{-q\phi_s/kT}\Bigl[\int_{0}^{1}\tau(\alpha)e^{-\frac{E_b}{kT}\alpha}d\alpha+e^{-\frac{E_b}{kT}}\Bigr].
\label{eq:R_1}
\end{eqnarray}

In Fig.2 is the contact resistivity obtained from simulations of the physical model for $N_D=1e20cm^{-3}$ and 
$T=300,500,700,900K$, respectively. We firstly compare it with results from a numerical model, namely model A. 
In model A, the parabolic potential variation is assumed by substituting Eqn.~\ref{eq:E_x} into Eqn.~\ref{eq:tau_PS}, which
gives
\begin{equation}
\tau^A(\alpha)=exp[-\frac{4}{3}\frac{E_b}{E_{00}}\sqrt{1-\alpha}\cdot (1-\sqrt{\alpha})].
\label{eq:tau_A}
\end{equation}
In this model, quasi-Fermi level is regarded as flat in silicon, i.e. Eqn.~\ref{eq:R_1} is used. It can be seen in Fig. 2 that
severe discrepancy of calculated resistivity exists between the numerical model A and the physical model
throughout the entire temperature range. In order to investigate its cause, we compare the potential variation
computed from these two models in Fig. 3. It is clearly observed that,
the assumption of fully depletion becomes invalid at energies near the quasi-Fermi level. Since the tunneling
probability exponentially depends on the tunneling distance, this discrepancy in $x-E$ relation leads to significant 
difference in the tunneling probability, as shown in Fig. 4. By removing the parabolic band-bending assumption in model A,
the accuracy of contact calculation can be greatly improved. For this purpose,
the one-dimensional Poisson equation needs to be solved for the depletion region. For M-B statistics, the Poisson equation 
is expressed as
\begin{eqnarray}
\frac{1}{q}\frac{d^2E}{dx^2}&=&\frac{q}{\epsilon}(N_D-n) \nonumber \\
&=&\frac{qN_D}{\epsilon}(1-e^{-\frac{E}{kT}}).
\label{eq:poisson}
\end{eqnarray}
Multiply $dE/dx$ on both sides of Eqn.~\ref{eq:poisson} and integrate from $0$ to $E$, and we obtain
\begin{equation}
\frac{dE}{dx}=\Bigl[\frac{2q^2N_D}{\epsilon}\int_0^E(1-e^{-\frac{E}{kT}})dE\Bigr]^{1/2}.
\end{equation}
Express it in terms of $\alpha$ and we have
\begin{equation}
F(\alpha)=\int_0^{\alpha}\Bigl(1-e^{-\frac{E_b}{kT}\alpha}\Bigr)d\alpha,
\label{eq:F_alpha_poisson}
\end{equation}
according to the definition of $F(\alpha)$ in Eqn.~\ref{eq:F_alpha}.
Eqn.~\ref{eq:F_alpha_poisson} therefore defines the potential variation from the exact solution to Poisson 
equation\footnote{After assuming constant quasi-Fermi level and ignoring hole concentrations.}. 
Apply Eqn.~\ref{eq:F_alpha} and~\ref{eq:F_alpha_poisson} to Eqn.~\ref{eq:tau_PS} and it is obtained that
\begin{equation}
\tau^B(\alpha)=exp\Bigl[-\frac{2}{3}\frac{E_b}{E_{00}}(1-\alpha)^{1/2}\int_{\alpha}^1\frac{d\alpha}
{\sqrt{F(\alpha)}}\Bigr].
\label{eq:tau_B}
\end{equation}
We then have an improved numerical model, namely model B, in which Eqn.~\ref{eq:F_alpha_poisson} and~\ref{eq:tau_B}
are used to compute the tunneling probability, and constant quasi-Fermi level (Eqn.~\ref{eq:R_1}) is still assumed.
We also plot the $x-E$ relation and the tunneling probability of model B in Fig.3 and Fig.4, respectively.
They show excellent match with those of the physical model. The computed resistivity from model B
is also plotted in Fig.2. It can be seen that qualitative improvement is obtained over model A with reference to the
physical model. At moderate temperature, the match between model B and the physical model is fairly good. However,
an evident discrepancy between them is still observed at high temperature, which is addressed in the next paragraph. 
As we previously mentioned, our numerical model A, B and the physical model are all based on Eqn.~\ref{eq:tau_PS},
which partially assumed triangular potential barrier {\it a priori}. Therefore, it would be interesting to see what
the effect of the potential variation is based on the the more general expression of tunneling 
probability (Eqn.~\ref{eq:tau_alpha}). As mentioned earlier, the expression used in~\cite{Crowell} 
(Eqn.~\ref{eq:tau_CR}) is derived from Eqn.~\ref{eq:tau_alpha} by assuming parabolic band-bending completely.
We use their expression Eqn.~\ref{eq:tau_CR} in numerical model $\tilde{A}$. In another numerical model $\tilde{B}$
we use the general expression Eqn.~\ref{eq:tau_alpha} and the realistic band-bending formula
Eqn.~\ref{eq:F_alpha_poisson}. The contact resistivity computed using these two models are plotted in Fig.5, and similar
trend is observed: the assumption of the parabolic band-bending severely under-estimates the resistivity at all temperature
range.

As shown in Fig.2, significant discrepancy in calculated resistivity exists at high temperature between numerical 
model B and the physical model. However, it can be seen from Fig.4 that the tunneling probability matches for the
two models. Therefore, the source of this discrepancy originates from the assumption of constant $\eta$ in the derivation
from Eqn.~\ref{eq:J_tot} to Eqn.~\ref{eq:J_tot_const}. In Fig.6 is the band-diagram of a Schottky contact simulated 
by the physical model for forward bias $V_f=0.1V$. The variation of the electron quasi-Fermi level within the depletion
region is evident due to finite carrier supply rate by drift-diffusion. If we let $V_f\rightarrow 0$, we expect this 
variation $\eta$ to approach zero at the same time. But
the ratio $\eta/V_f$ can be finite in this limit. Since the resistivity is a differential quantity, it can be affected
by this effect. However, it should be noted that the validity of drift-diffusion model in the
depletion region is an open question itself, since the depletion width is comparable to electron mean free path. If the carrier
supply is not limited by the drift-diffusion process, the variation of quasi-Fermi level can be negligible. A possible
way to examine this problem is Monte Carlo simulation. The electron continuity equation is
\begin{equation}
\frac{dj}{dx}+U=0,
\label{eq:con}
\end{equation}
where $j$ is electron flux due to carrier transport in the depletion region, and $U$ is tunneling flux density.
Plug in the tunneling probability and integrate from $0$ to $w$ and we obtain that
\begin{eqnarray}
j(\alpha)&=&\frac{AE_b}{qkT}e^{-\frac{q\phi_s}{kT}}\Bigl[\int_0^{\alpha}\tau(\alpha')e^{-\frac{E_b}{kT}\alpha'}
(e^{-\frac{q\eta(\alpha')}{kT}}-e^{-\frac{qV_f}{kT}})d\alpha'+e^{-\frac{E_b}{kT}}(e^{-\frac{q\eta_b}{kT}}-
e^{-\frac{qV_f}{kT}})\Bigr] \\
&\approx&\frac{AE_b}{qkT}e^{-\frac{q\phi_s}{kT}}(e^{-\frac{q\eta}{kT}}-e^{-\frac{qV_f}{kT}})
\Bigl[\int_0^{\alpha}\tau(\alpha')e^{-\frac{E_b}{kT}\alpha'}d\alpha'\Bigr] \\
&\equiv& \frac{A}{q}e^{-\frac{q\phi_s}{kT}}(e^{-\frac{q\eta}{kT}}-e^{-\frac{qV_f}{kT}})P(\alpha).
\label{eq:j_con}
\end{eqnarray}
If we assume the electron transport in the depletion region can be modeled as drift-diffusion, as in the physical model,
the electron flux can also be expressed as
\begin{eqnarray}
j(\alpha)&=&\mu N_De^{-\frac{E_b}{kT}\alpha}e^{-\frac{q\eta}{kT}}\frac{d\eta}{dx} \nonumber \\
&=& \mu N_De^{-\frac{E_b}{kT}\alpha}e^{-\frac{q\eta}{kT}}\sqrt{\frac{2q^2N_D}{\epsilon E_b}F(\alpha)}\cdot \frac{d\eta}{d\alpha},
\label{eq:j_dd}
\end{eqnarray}
where $\mu$ is electron mobility.
Compare Eqn.~\ref{eq:j_con} and~\ref{eq:j_dd} and let $V_f,\eta \rightarrow 0$, then we obtain
\begin{equation}
\frac{dlog(1-\eta/V_f)}{d\alpha}=-\frac{A}{kTq\mu}e^{-\frac{q\phi_s}{kT}}\sqrt{\frac{\epsilon E_b}{2N_D^3}} \cdot
\frac{P(\alpha)}{e^{-\frac{E_b}{kT}\alpha}\sqrt{F(\alpha)}},
\end{equation}
or equivalently,
\begin{equation}
1-\eta/V_f=exp\Bigl[-\int_0^{\alpha}-\frac{A}{kTq\mu}e^{-\frac{q\phi_s}{kT}}\sqrt{\frac{\epsilon E_b}{2N_D^3}} \cdot
\frac{P(\alpha')}{e^{-\frac{E_b}{kT}\alpha'}\sqrt{F(\alpha')}}d\alpha'\Bigr].
\end{equation}
The inverse of resistivity near zero bias is then revised as
\begin{equation}
R^{-1}=\frac{AqE_b}{(kT)^2}e^{-\frac{q\phi_s}{kT}}\Bigl[\int_0^1\tau(\alpha)e^{-\frac{E_b}{kT}\alpha}(1-\eta/V_f)d\alpha
+e^{-\frac{E_b}{kT}}\Bigr].
\end{equation}
The resistivity calculated using this improved model (model C) is also plotted in Fig.2, and excellent match with the results
of the physical model is observed.

\clearpage

\begin{figure}
\begin{center}
\begin{tabular}{c}
\includegraphics[height=10cm]{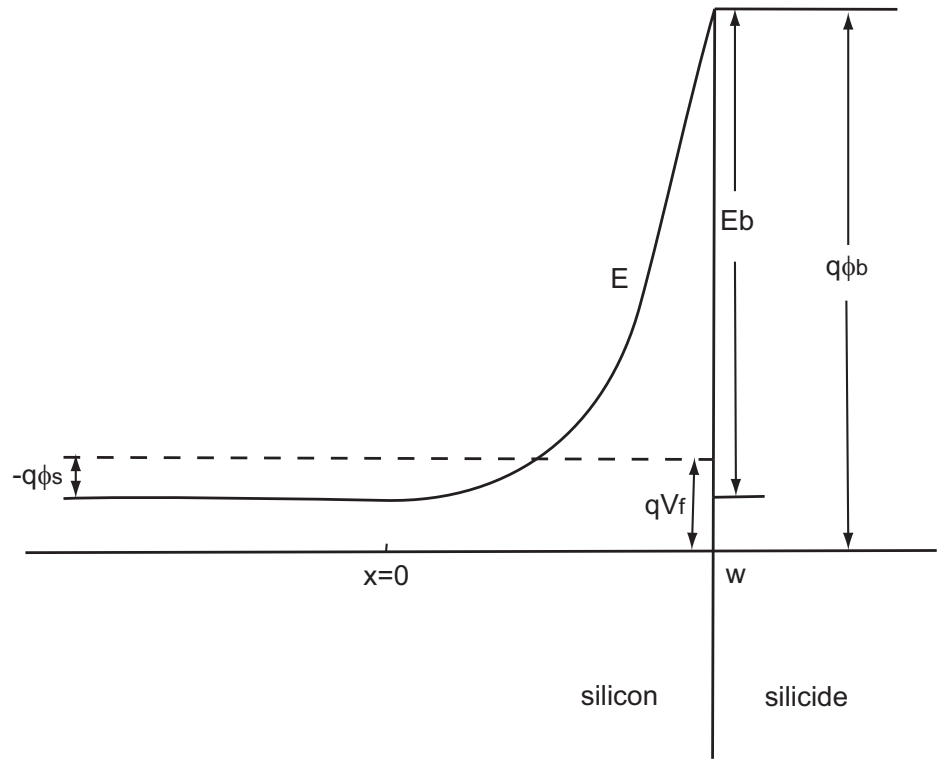}
\end{tabular}
\end{center}
\caption{ \label{fig_1}
A schematic plot of energy diagram of a silicon/silicide Schottky contact.
 }
\end{figure}

\begin{figure}
\begin{center}
\begin{tabular}{c}
\includegraphics[height=10cm]{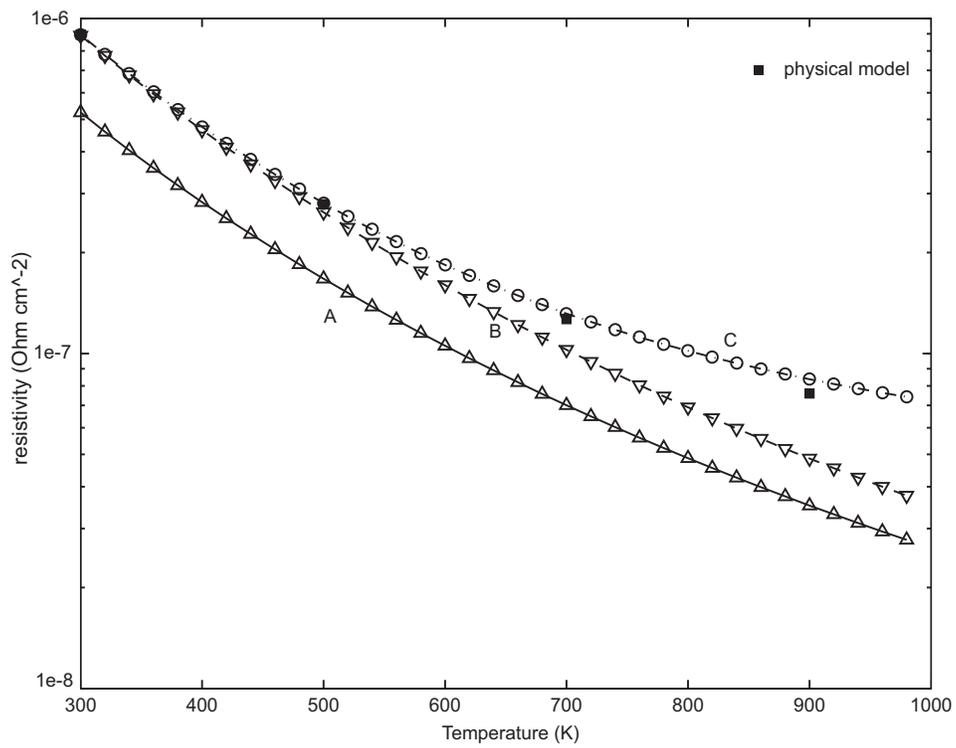}
\end{tabular}
\end{center}
\caption{ \label{fig_2}
Contact resistivity vs. temperature calculated by various numerical models and the physical model.
 }
\end{figure}

\begin{figure}
\begin{center}
\begin{tabular}{c}
\includegraphics[height=10cm]{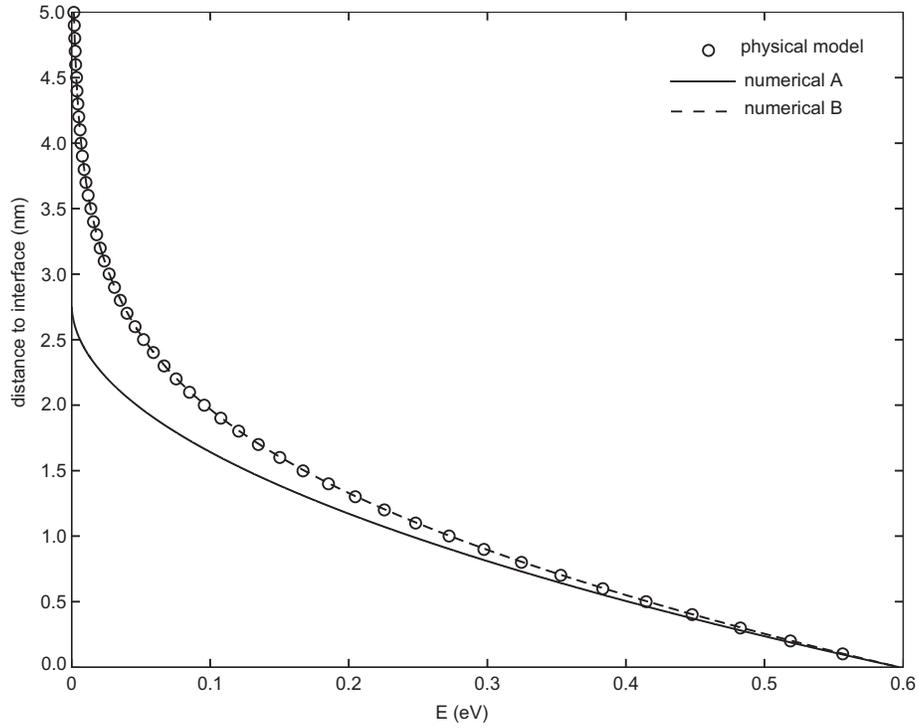}
\end{tabular}
\end{center}
\caption{ \label{fig_3}
Distance to the interface vs. barrier energy (x-E relation) calculated by various numerical models and the physical model.
 }
\end{figure}

\begin{figure}
\begin{center}
\begin{tabular}{c}
\includegraphics[height=10cm]{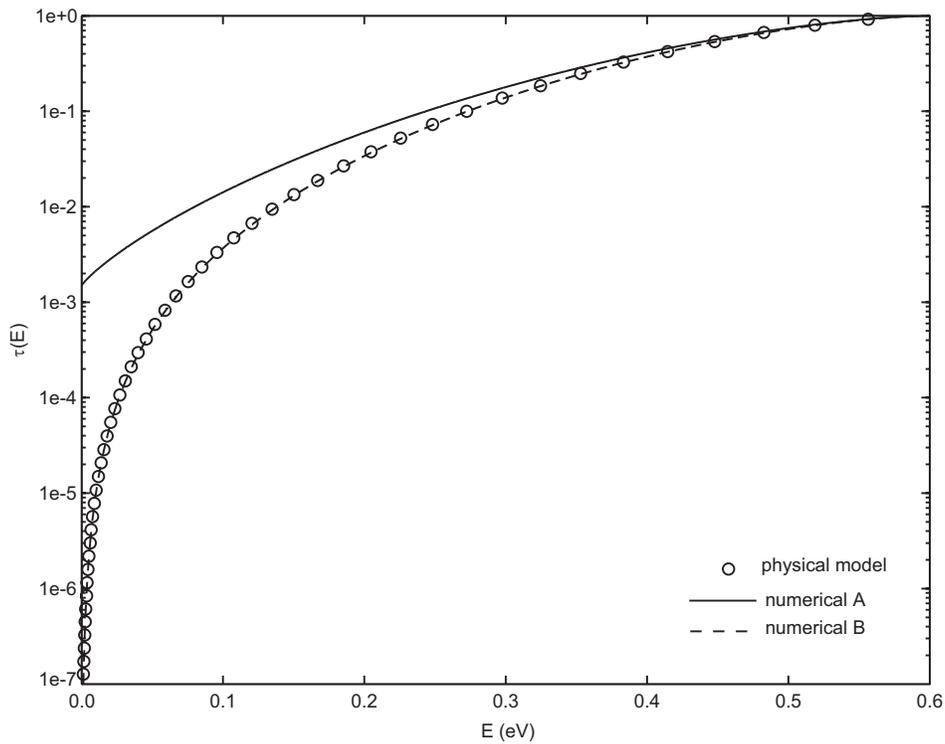}
\end{tabular}
\end{center}
\caption{ \label{fig_4}
Tunneling probability vs. barrier energy calculated by various numerical models based on Eqn.~\ref{eq:tau_PS} 
and the physical model.
 }
\end{figure}

\begin{figure}
\begin{center}
\begin{tabular}{c}
\includegraphics[height=10cm]{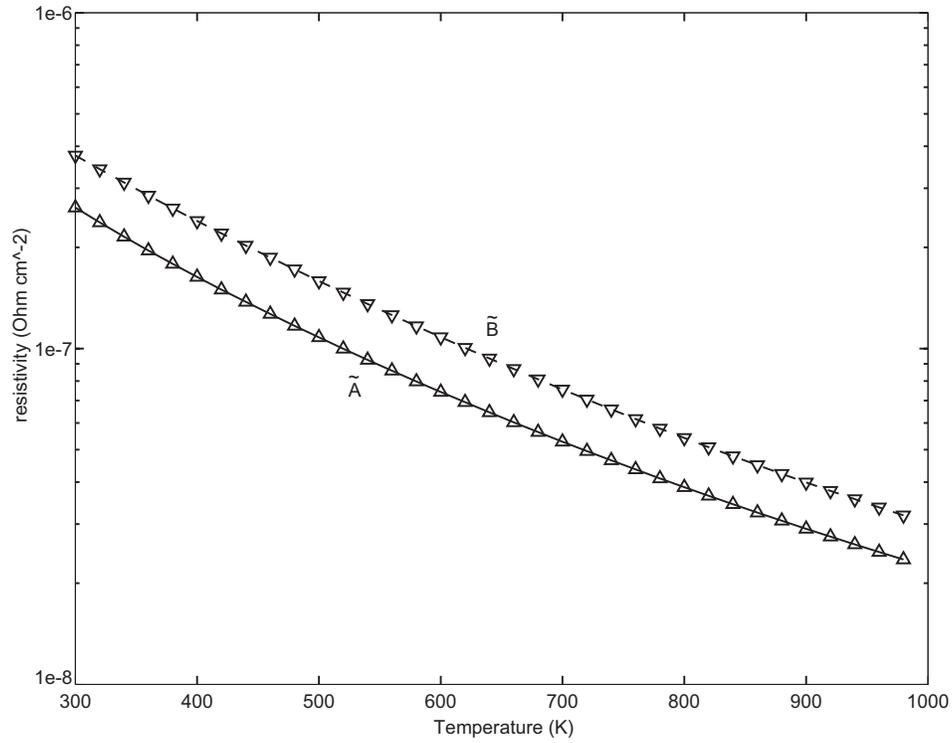}
\end{tabular}
\end{center}
\caption{ \label{fig_5}
Contact resistivity calculated by two numerical models based on Eqn.~\ref{eq:tau_alpha}.
}
\end{figure}

\begin{figure}
\begin{center}
\begin{tabular}{c}
\includegraphics[height=10cm]{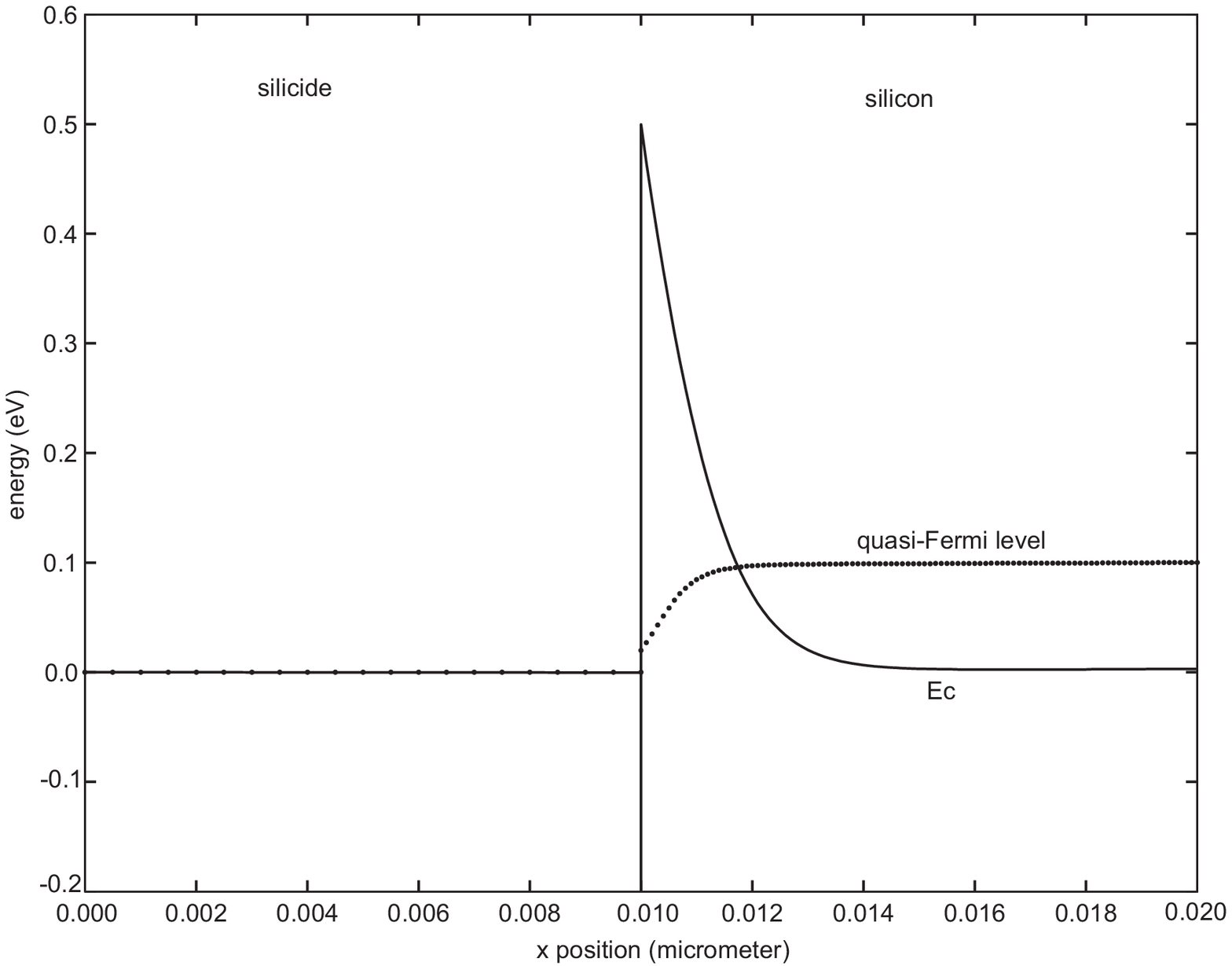}
\end{tabular}
\end{center}
\caption{ \label{fig_6}
Energy band diagram of a Schottky contact at $0.1V$ forward bias simulated by the physical model. 
 }
\end{figure}


\begin{thebibliography}{10}
\bibitem{Hu}
M.-C. Jeng, J. E. Chung, P.-K. Ko, and C. Hu, IEEE Trans. Elec. Dev. {\bf 37}, 2408 (1990).

\bibitem{Padovani}
F. A. Padovani and R. Stratton, Solid-State Electronics {\bf 9}, 695 (1966).

\bibitem{Crowell}
C. R. Crowell and V. L. Rideout, Solid-State Electronics {\bf 12}, 89 (1969).

\bibitem{Yu}
A. Y. C. Yu, Solid-State Electronics {\bf 13}, 239 (1970).

\bibitem{Matsuzawa}
K. Matsuzawa, K. Uchida, and A. Nishiyama, IEEE Trans. Elec. Dev. {\bf 47}, 103 (2000).

\bibitem{Bardeen}
J. Bardeen, Phys. Rev. Lett. {\bf 6}, 57 (1961).

\bibitem{Harrison}
W. A. Harrison, Physcal Review {\bf 123}, 85 (1961).

\end{thebibliography}
\end{document}